\documentclass[twoside,12pt]{article}
\usepackage{graphicx}
\usepackage{amssymb}
\usepackage{a4}

\begin{document}                                                                
       
\title{Dirichlet Boundary Value Problems of the Ernst Equation}
\author{Marcus Ansorg, Andreas Kleinw\"{a}chter,\\ Reinhard
Meinel, Gernot Neugebauer\\[5mm]
{\it Theoretisch-Physikalisches Institut, University of Jena,}\\
{\it Max-Wien-Platz 1, 07743 Jena, Germany}}
\date{}
\maketitle

\begin{abstract}\noindent
We demonstrate how the solution to an exterior Dirichlet boundary
value problem of the axisymmetric, stationary Einstein equations
can be found in terms of generalized solutions of the
B\"{a}cklund type. The proof that this generalization
procedure is valid is given, which also proves conjectures 
about earlier representations of the gravitational field corresponding 
to rotating disks of dust in terms of B\"{a}cklund type solutions.
As a further result, we find that in contrast to the Laplace
equation, arbitrary boundary values may not be prescribed.
\end{abstract}

\section{Introduction}
Extraordinarily massive, compact astrophysical objects like neutron
stars require a fully relativistic treatment. This motivates the
study of the relativistic gravitational field of axisymmetric,
stationary rotating bodies. \\
The investigation of the exterior field, i.e.~of arbitrary
exterior boundary value problems of the axisymmetric stationary
Einstein equations, can only be a first step in this direction, for
neither the shape of the body (i.e.~the boundary) nor particular
boundary values may be prescribed. For a given equation of state
characterizing the matter of the rotating body, the boundary and
the values there result from transition conditions since the
metric along with its first normal derivative behave
continuously at the boundary (in appropriate coordinates).
Therefore, a subsequent step must be the combination of a
procedure to compute the interior solution with the method to
treat the exterior field
in order to realize the transition conditions. \\
Our aim is to show a way to handle the exterior
field. In particular, we want to make use of the variety of exact
analytic solutions that are available in the exterior region.
Thanks to the facts that
\begin{enumerate}
  \item the exterior vacuum field equations can be summarized in
        a single nonlinear complex differential equation -- the
        so-called Ernst equation\cite{Ernst,Ernsta},
  \item the Ernst equation has been found to be the integrability
        condition of a corresponding linear matrix problem,
\end{enumerate}
it is possible to apply various soliton methods (see
\cite{Mais78}-\cite{Hauserc}) to create explicit exterior
solutions, among them the B\"{a}cklund transformations and the
Riemann-Hilbert techniques. By extensive investigation of the
latter, Neugebauer and Meinel succeeded in solving the
boundary value problem that corresponds to an infinitesimal thin
rigidly rotating disk of dust\cite{NM93,NM95,NKM96}. In
particular, they found their solution to belong to 
a `hyperelliptic class' of 
solutions\cite{MN96}, see also \cite{Kor89}-\cite{MN97}. Further
investigations as to whether this class of solutions can be used
to solve more general boundary value problems corresponding to
differentially rotating disks of dust have been carried out by
Ansorg and Meinel\cite{AM00} and by Ansorg\cite{A01}. In the
latter paper it was demonstrated that the B\"{a}cklund
type solutions \footnote{The B\"{a}cklund type solutions can be
obtained by applying B\"{a}cklund transformations to purely real
`seed'-solutions of the Weyl class. These solutions form a
subclass of the hyperelliptic class, see\cite{A01}.} suffice to approximate 
the exterior
gravitational field of an arbitrary differentially rotating disk
of dust.\\
In this paper we investigate the question, as to whether the
treatment of the boundary value problem of an arbitrary
differentially rotating disk of dust by means of B\"{a}cklund
type solutions can also be applied to extended boundaries $B$, 
which can potentially form the shape of a rotating body. We denote
by $B$ a smooth spatial curve, at which regular boundary values
are prescribed. The solubility of such 
Dirichlet boundary value problems for 
sufficiently weak relativistic boundary values has been proven 
by Reula\cite{Reula}, see also \cite{Pfister,Schaudt}.
In illustrative examples we will take spherical
boundaries in order to make the formulas more
transparent and therefore get better insight into the
underlying mathematical structure. However, the general statements are
valid for arbitrary extended $B$.\\
The paper is organized as follows. At first, the metric tensor
and the Ernst equation are introduced. Then, solution techniques
to solve boundary value problems of the real axisymmetric 
three-dimensional Laplace equation with
the boundary $B$ are discussed in order to prepare the
relativistic treatment. These solutions will be given in terms of
an analytic function $H$, which is defined on a curve $\Gamma$
in the complex plane and satisfies $H(\overline{X})=\overline{H(X)}$
(with $X\in\mathbb{C}$ and the bar denoting complex conjugation).
Here, the curve $\Gamma$ is closely connected with the boundary $B$.
In this formulation,
the solutions describe both a regular interior and exterior
field, which (in general) assumes different values on the inner
and outer side of $B$. The prescription of
these interior and exterior boundary values uniquely determines
the function $H$, i.e.~this formulation permits the simultaneous
solution of an interior and exterior boundary value problem.\\
The second section treats the hyperelliptic solutions of the
Ernst equation and their generalization by means of a suitable
limiting process. Similar to the formulation of solutions of the
real Laplace equation, we find these generalized solutions
\begin{enumerate}
  \item to depend on an analytic function $\gamma$ defined on $\Gamma$
  (and, in general, not subject to further
  requirements)
  \item to permit the simultaneous solution of an
  interior and exterior Dirichlet boundary value problem of the Ernst
  equation.
\end{enumerate}
In the case of a flat interior field we find an
explicit relation of the function $\gamma$ and the values
of the Ernst potential at the exterior part of the symmetry axis. \\
In the section 3.1, we investigate the solutions of the
B\"{a}cklund type. Apart from regularity properties we discuss the
generalization of these solutions which can be performed since
they form a subclass of the hyperelliptic solutions. As a result
we prove conjectures formulated in \cite{A01}, in particular that the
Neugebauer-Meinel-solution can be written as a well defined
limit of B\"{a}cklund type solutions.\\
Because of their mathematical simplicity (compared with the much
more complicated hyperelliptic solutions), the B\"{a}cklund type
solutions play the fundamental role in our treatment for
approximating solutions of an exterior boundary value
problem by analytic solutions. This is carried out in section
3.2. As a result, we find that in contrast to the Laplace
equation, arbitrary boundary values of the Dirichlet type may not
be prescribed.\\
In what follows, units are used in which the velocity of light as
well as Newton's constant of gravitation are equal to 1.
\subsection{Metric tensor, Ernst equation, and boundary
conditions} The metric tensor for axisymmetric stationary and
asymptotically flat vacuum space-times reads as follows in
Weyl-Papapetrou-coordinates $(\rho,\zeta,\varphi,t)$:
\[ds^2=e^{-2U}[e^{2k}(d\rho^2+d\zeta^2)+\rho^2d\varphi^2]
-e^{2U}(dt+a\;d\varphi)^2\,.\] 
The field equations are equivalent to a single complex equation --
the so-called Ernst equation 
   \begin{equation}
   \label{G1.1}
   (\Re f)\; \triangle f=({\bf \nabla} f)^2\,,
   \end{equation}
   \[\triangle=\frac{\partial^2}{\partial\rho^2}+\frac{1}{\rho}
   \frac{\partial}{\partial\rho}+\frac{\partial^2}{\partial\zeta^2},
   \qquad \nabla=\left(\frac{\partial}{\partial\rho},\frac{\partial}
   {\partial\zeta}\right),\]
   where the Ernst potential $f$ is given by
   \begin{equation}
   \label{G1.2}
   f=e^{2U}+\mbox{i}\,b\quad\mbox{with}\quad b_{,\zeta}=\frac{e^{4U}}
   {\rho}\,a_{,\rho}\;,\quad
   b_{,\rho}=-\frac{e^{4U}}{\rho}\,a_{,\zeta}.
   \end{equation}
The remaining function $k$ can be calculated from the Ernst
potential $f$ by a line integral:
\begin{eqnarray*}
   \frac{k_{,\rho}}{\rho}&=&(U_{,\rho})^2-(U_{,\zeta})^2
   +\frac{1}{4}e^{-4U}[(b_{,\rho})^2-(b_{,\zeta})^2]\;, \\
   \frac{k_{,\zeta}}{\rho}&=&2U_{,\rho} U_{,\zeta}
   +\frac{1}{2}e^{-4U}b_{,\rho} b_{,\zeta}
   \end{eqnarray*}
In this paper we treat the following boundary value problem of the
Ernst equation (\ref{G1.1}):
\begin{enumerate}
  \item The boundary $B$ is given by a smooth spatial curve which can
        be described by some positive analytic function
        $r_B:[-1,1]\to\mathbb{R}_+$:
        \[B=\{(\rho,\zeta):\rho=r_B(\cos\vartheta)\sin\vartheta,\,
        \zeta=r_B(\cos\vartheta)\cos\vartheta,\,\vartheta\in[0,\pi]\}\]
  \item Along the boundary $B$ we require regular
        boundary conditions of the Dirichlet type,
        i.e.~$f$ is given for $(\rho,\zeta)\in B$.
  \item Regularity at the rotation axis $\rho=0$ is guaranteed by
      \[\frac{\partial f}{\partial\rho}(0,\zeta)=0.\]
  \item At infinity, asymptotic flatness is realized by $f\to 1$.
\end{enumerate}
The treatment in section 2 will provide us a formulation
$f=f(\gamma)$ which is sufficiently general to satisfy interior
and exterior boundary conditions of the Ernst equation
(\ref{G1.1}), i.e.~for an appropriate prescription\footnote{As
already mentioned above, the Dirichlet boundary values
cannot be chosen arbitrarily.}  of (not necessarily coinciding) interior and
exterior boundary values, there is a uniquely determined function
$\gamma:\Gamma\to\mathbb{C}$ [with $\Gamma=\{X\in\mathbb{C}:(|\Im
X|\,,\Re X)\in B\}$] such that the Ernst potential $f=f(\gamma)$
is both regular within and without $B$ and assumes the above
boundary values at the inner and outer side of $B$.
Since for physical reasons we are only interested in an exterior
solution, the freedom in the choice of the interior boundary
values will be used to restrict the function $\gamma$ such that
the solution $f=f(\gamma)$ can be represented by B\"{a}cklund
type solutions. This is outlined in the third section.
Furthermore, for the exterior solution we will impose the additional 
physical requirement
of reflectional symmetry with respect to the plane $\zeta=0$ (see \cite{MN95})
which leads to
\[\begin{array}{llll}r_B(\tau)&=&r_B(-\tau)&\mbox{for all $\tau\in[-1,1]$}\,,\\[3mm]
                    f(\rho,-\zeta)&=&\overline{f(\rho,\zeta)}
&\mbox{for all $(\rho,\zeta)$ outside $B$.}\end{array}\]
\subsection{Boundary value problems of the Laplace equation}
In this section we prepare the general formulation $f=f(\gamma)$
by considering the corresponding non-relativistic gravitational
boundary value problem -- i.e.~the general boundary value
problem of the axisymmetric three-dimensional 
Laplace equation \[\triangle U=0.\] Any real
solution $U$ which is is both regular inside and outside the
boundary $B$ can be written in the form
\begin{equation}
\label{Lapl}
  U(\rho,\zeta;H)=\frac{1}{2\pi\mbox{i}}\oint_\Gamma
  \frac{H(X)dX}{W_z}
\end{equation}
with
\begin{eqnarray}\nonumber
&&\Gamma=\{X\in\mathbb{C}:(|\Im X|\,,\Re X)\in B\}\,,\\
&&\nonumber\mbox{$H:\Gamma\to\mathbb{C}$ is an analytic function with}\quad
H(\overline{X})=\overline{H(X)}\,,\\
\label{DefWz}
&&  W_z=\sqrt{(X-\zeta)^2+\rho^2},\\
&&\quad\nonumber\Re(W_z)=\left\{\begin{array}{lcl} <0\quad \mbox{for
$(\rho,\zeta)$ outside
$B$}& \mbox{or}
                                      & \Re(X)<\zeta \\
                                   >0\quad\mbox{for $(\rho,\zeta)$ inside $B$}&
                                   \mbox{and}
                                       &   \Re(X)>\zeta
                  \end{array}\right. .
\end{eqnarray}
The function $H$ is uniquely determined by the interior and
exterior boundary values of the potential $U$. For a spherical
boundary (with radius $r_B\equiv 1$) we may write $H$ in the form
\[H(X)=\sum_{j=1}^\infty (H_j^{(+)} X^{j-1}+H_j^{(-)}X^{-j})
\,,\quad H_j^{(\pm)}\in\mathbb{R}.\] Then the coefficients
$H_j^{(\pm)}$ can be read from an expansion of the boundary
values in Legendre polynomials
\[P_n(x)=\frac{1}{2^n n!}\frac{d^n}{dx^n}[(x^2-1)^n]\]
of the cosine of
the angular coordinate $\vartheta$ with
$\rho=r\sin\vartheta,\,\zeta=r\cos\vartheta$:
\begin{eqnarray*}\lim_{r\uparrow 1} U(\rho,\zeta;H)&=&\sum_{j=1}^\infty
H_j^{(+)}P_{j-1}(\cos\vartheta)\\
\lim_{r\downarrow 1}
U(\rho,\zeta;H)&=&-\sum_{j=1}^\infty
H_j^{(-)}P_{j-1}(\cos\vartheta)\end{eqnarray*}
\subsubsection*{Discussion.}
\begin{enumerate}
  \item Reflectional symmetry
    \begin{equation}\label{symN}U(\rho,-\zeta;H)=\left\{\begin{array}{lll}
       \kappa U(\rho,\zeta;H) &\mbox{for}& \mbox{$(\rho,\zeta)$ within $B$}\\
      -\kappa U(\rho,\zeta;H)&\mbox{for}& \mbox{$(\rho,\zeta)$ without $B$}
      \end{array}\right.,\end{equation}
    with $\kappa^2=1$ is obtained if $r_B(\tau)=r_B(-\tau)$ and 
    $H(-X)=\kappa H(X)$, i.e.~an
    odd/even function $H$ produces an even/odd exterior
    potential $U$ in the coordinate $\zeta$.
  \item The freedom of the choice of the interior field can be used to
   restrict the function $H$ in an appropriate manner. For example we may choose 
   $\Re(H)=\mbox{const.}$ for $X\in\Gamma$, where this constant can be arbitrarily prescribed.
   For the above spherical case, this results in
   $H_{j+1}^{(+)}=-H_j^{(-)}$.
   Then the function $H$ is uniquely determined only by the exterior boundary values.
   Analogously, in the relativistic treatment (see section 3.2) we may choose the
   corresponding function $\gamma$ such that $\Im(\gamma)=\mbox{const.}\neq 0$ for $X\in\Gamma$, 
   which ensures that the
   corresponding Ernst potential $f$ can be approximated very well by solutions of the B\"{a}cklund
   type.
\end{enumerate}

\section{The hyperelliptic class of solutions and its generalization
by a suitable limiting process}
Meinel and Neugebauer \cite{MN96} as well as Korotkin \cite{Kor89,Kor93},
see also \cite{Kor97,MN97}, were able to construct the hyperelliptic class
of solutions containing a finite number of complex parameters and one
arbitrary real solution to the axisymmetric three-dimensional Laplace
equation. By investigation of corresponding Riemann-Hilbert-problems
one finds, that this class can be generalized in an appropriate manner.
The generalized class turns out to contain all Ernst
potentials which are both regular within and without the boundary $B$, which
assume (in general) different boundary values at the inner and outer side
of $B$, and are sufficiently weakly relativistic, i.e.~close
to the flat space solution $f\equiv1$.
In this section we discuss these generalized solutions which allow us to
approximate the solution to a simultaneous interior and exterior boundary
value problem.
\subsection{The hyperelliptic class of solutions}
We adapt the hyperelliptic class of solutions as given in
\cite{MN96} to our purposes by writing:\\
For a given integer $p\geq 1$, a set
$\{X_1,\ldots,X_p\}=\{X_\nu\}_p$\footnote{In the following, the
notation $\{X_1,\ldots,X_p\}$ will be abbreviated by
$\{X_\nu\}_p$.} of complex parameters, and an analytic function
$H:\Gamma\to\mathbb{C},\,H(\overline{X})=\overline{H(X)}$, the
following expression
\begin{eqnarray}
    \label{hyper}
    f(\rho,\zeta;\{X_\nu\}_p\,,H)=
    \exp\left(\sum_{\nu=1}^p\int_{X_\nu}^{X^{(\nu)}}
    \frac{X^p\,dX}{W}+\frac{1}{2\pi\mbox{i}}\oint_\Gamma
      \frac{H(X)X^p\,dX}{W_z}\right)
\end{eqnarray}
 with
\begin{eqnarray}\nonumber
&&W=\sqrt{(X+\mbox{i}z)(X-\mbox{i}\bar{z})\prod_{\nu=1}^p(X-X_\nu)(X-\bar{X}_\nu)}\,,\quad
   z=\rho+\mbox{i}\zeta,\\
\label{JUP}&&
\sum_{\nu=1}^p\int_{X_\nu}^{X^{(\nu)}}\frac{X^j\,dX}{W}=-\frac{1}{2\pi
\mbox{i}}\oint_\Gamma \frac{H(X)X^j\,dX}{W_z},   \quad 0\leq
j<p.\end{eqnarray} satisfies the Ernst equation. The
($z$-dependent) values for the $X^{(\nu)}$ as well as the
integration paths on a two-sheeted Riemann surface have to be
taken from the solution to the  Jacobian inversion problem
(\ref{JUP}). \\
The Ernst potential $f=f(\{X_\nu\}_p\,,H)$ \footnote{For
simplicity, we suppress the coordinate dependence and write only
$f=f(\{X_\nu\}_p\,,H)$.} is regular at
$(\rho,\zeta)=(|\Im[X_\nu]|,\Re[X_\nu])$ as can be deduced from an
appropriate combination of the equations (\ref{JUP}) and
corresponding rearrangements of the terms occurring in (\ref{hyper}).
In general, $f(\{X_\nu\}_p\,,H)$ assumes different values at the
inner and outer sides of the boundary $B$, i.e.~it possesses a
jump along $B$.
\subsection{The generalization of the hyperelliptic class of
solutions by a suitable limiting process}
This section contains a theorem which ensures that the above hyperelliptic
solutions can be generalized in an appropriate manner such that simultaneous
interior and exterior boundary value problems become soluble by means of 
these solutions.
The theorem consists of three parts, the proofs of which are outlined.
Moreover, we illustrate certain properties of the generalized class.
\subsubsection*{Theorem.}
Given the analytic function $\gamma:\Gamma\to\mathbb{C}$
which can be extended to some compact neighborhood
$\mathbb{G}_\gamma$  about $\Gamma$, then
\begin{enumerate}
\item for sufficiently small $\varepsilon$, the Ernst potential
\[f(\varepsilon\gamma)=\lim_{p\to\infty}f(\{X_\nu^{(p)}\}_p\,,\varepsilon H_p)\]
exists and is independent of the particular choice of the
sequences $\{\{X_\nu^{(p)}\}_p\}_{p_0}^\infty$ and
$\{H_p\}_{p_0}^\infty$ which serve to represent $\gamma$ by
\begin{eqnarray}\label{chigen}
\gamma(X)=\lim_{p\to\infty}\,H_p(X)\prod_{\nu=1}^p(X-X^{(p)}_\nu)\quad
\mbox{for}\quad X\in\mathbb{G_\gamma}.    \nonumber
\end{eqnarray}
\item the Ernst potential $f=f(\varepsilon\gamma)$ is  both regular inside
and outside the boundary $B$ and assumes (in general) different boundary
values at the inner and outer side of $B$. The exterior and interior
Ernst potentials can be extended beyond the boundary $B$ to the region
$\mathbb{B}_\gamma=\{(\rho,\zeta):\zeta\pm\mbox{i}\rho\in\mathbb{G_\gamma}\}$.
\item any sufficiently weak relativistic Ernst potential $f$ which is both
regular within and without the boundary $B$ with (in general) different boundary
values at the inner and outer side of $B$, can be written in the
above manner as $f=f(\gamma)$. The function $\gamma$ is uniquely determined
by the Dirichlet boundary values of $f$ at $B$.
\end{enumerate}  
\subsubsection*{Proof.}
\begin{enumerate}
\item In order to prove the first one of the above statements we establish a relation
between the function $\gamma$ and holomorphic functions $\alpha$
and $\beta$ defined on $\mathbb{G}_\gamma$ from which the Ernst
potential $f$ can be calculated via the solution of a
Riemann-Hilbert-Problem. 
\begin{enumerate}
\item To introduce $\alpha$ and $\beta$, we
follow the treatments of \cite{NM93}, \cite{N96} and \cite{N00}: \\
The Ernst equation is the integrability condition of the Linear Problem
\begin{equation}\label{LS}\mbox{\normalsize $\begin{array}{lll}
\Phi_{,z}&=&[f+\bar{f}\,]^{-1}\left(\begin{array}{cc}
\bar{f}_{,z}& \lambda\bar{f}_{,z}\\
\lambda f_{,z}& f_{,z}\end{array}\right)\Phi\\[5mm]
\Phi_{,\bar{z}}&=&[f+\bar{f}\,]^{-1}\left(\begin{array}{cc} \bar{f}_{,\bar{z}}&
\lambda^{-1}\bar{f}_{,\bar{z}}\\[1mm]
\lambda^{-1} f_{,\bar{z}}& f_{,\bar{z}}\end{array}\right)\Phi
\end{array}$}
\end{equation}
where $\Phi=\Phi(X,\rho,\zeta)$ is a $2\times 2$-matrix and the
spectral parameter $\lambda$ is given by
\[\lambda(X)=\sqrt{\frac{X-\mbox{i}\bar{z}}{X+\mbox{i}z}}\,.\]
One explicit way to create solutions to the Ernst equation that
are regular within and without $B$ and possess a jump at $B$ is to use
the Riemann-Hilbert-techniques, by which we force the
corresponding matrix $\Phi$ to possess a multiplicative jump at
$\Gamma$:
\[\Phi_+=\Phi_-C(X)\] and moreover to be regular within
and without $\Gamma$. The indices $'+'$ and $'-'$ refer to the
interior and exterior sides of $\Gamma$, respectively. The
jump-matrix $C=C(X)$ is independent of the coordinates
$(z,\bar{z})$ and can be cast into the form
\[C(X)=\left(\begin{array}{cc} \alpha(X)& 0\\ \beta(X)&1\end{array}\right), \quad
\overline{\alpha(X)}=\alpha(\overline{X})\,,\quad\overline{\beta(X)}=-\beta(\overline{X}).\]
If the functions $\alpha$ and $\beta$ are prescribed, the Ernst potential at an arbitrary point
$(\rho,\zeta)$ can be determined by the solution of a linear integral equation, see \cite{N96}.\\
As outlined in \cite{N00}, the linear system can be integrated along the
rotation axis, which yields in agreement with the above Rieman-Hilbert-Problem
[$\rho=0$, $\zeta>-r_B(-1)$]:
\begin{equation}\label{Pax}\mbox{\normalsize $\begin{array}{lll}
\Phi^+&=&\left(\begin{array}{cc} \bar{f}& 1\\
f&-1\end{array}\right) \left(\begin{array}{cc} F(X)& 0\\
B(X)&1\end{array}\right)\\[5mm]
\Phi^-&=&\left(\begin{array}{cc} 1&\bar{f}\\
1&-f\end{array}\right) \left(\begin{array}{cc} 1&B(X)\\0&
F(X)\end{array}\right)\end{array}$}
\end{equation} The notation $\Phi^\pm$
refers to the two sheets on which $\Phi$ is defined; $'\pm'$ means
the sheet in which $\lambda=\pm 1$ for $\rho=0$. The functions
$F=F(X)$ and $B=B(X)$ 
\footnote{The function $B=B(X)$ is not to be confused with the boundary $B$.}
are defined in the complex plane and are regular
within and without the contour $\Gamma$ ($F\to 1$ and $B\to 0$ as
$|X|\to\infty$) but possess a jump at $\Gamma$. This jump can be
taken from the above Riemann-Hilbert-Problem:
\begin{equation}\label{RHP}F_+=\alpha F_-\,,\quad B_+=\alpha B_-+\beta.\end{equation}
\item We establish a connection between the relativistic parameter
$\varepsilon$ as well as $\gamma$
and the functions $F_\pm$ and $B_\pm$, now additionally depending on $\varepsilon$.
Thus, by virtue of (\ref{RHP}), we get
 $\alpha=\alpha(X;\varepsilon;\gamma)$ and $\beta=\beta(X;\varepsilon;\gamma)$. \\[5mm]
Given the function $\gamma$ which is holomorphic on
$\mathbb{G}_\gamma$, we denote by $\cal{B}_\gamma$ the Banach
space of all holomorphic functions on $\mathbb{G}_\gamma$ with
the norm
\[||\,\sigma||=\sup_{X\in\,\mathbb{G}_\gamma}|\,\sigma(X)|\quad\mbox{for}\quad
\sigma\in\cal{B}_\gamma.\] We now define functions
\[ F_\pm:[0,\varepsilon_\gamma]\to{\cal B}_\gamma\,,\quad
B_\pm:[0,\varepsilon_\gamma]\to {\cal B}_\gamma\] that satisfy
the following system of differential equations:
\begin{equation}\label{Dgl}\mbox{\normalsize $\begin{array}{lll}
\varepsilon \mbox{\Large $\frac{d}{d\varepsilon}$}
F_\pm&=&F_\pm(\hat{L}_\pm
s_1)+B_\pm(\hat{L}_\pm s_2)\\[5mm]
2\varepsilon \mbox{\Large $\frac{d}{d\varepsilon}$}
B_\pm&=&\mbox{\Large $\frac{B_\pm^2-1}{F_\pm}$}(\hat{L}_\pm s_2)
+F_\pm(\hat{L}_\pm s_3).\end{array}$}
\end{equation}
Here, the $s_j$ as well as the $\hat{L}_\pm$ are introduced as
follows:
\begin{enumerate}
  \item The $s_j$, $s_j:[0,\varepsilon_\gamma]\to{\cal B}_\gamma$,
  follow from $F_\pm, B_\pm$ by\\[1mm]
  \[\begin{array}{l}
    [(F_++F_-)s_1+(B_++B_-)s_2]\tanh\frac{G}{2}=G(F_--F_+)\\[5mm]
    \left[(B_++B_-)s_1+\left(\frac{B_+^2-1}{F_+}+\frac{B_-^2-1}{F_-}\right)
    s_2\right]\tanh\frac{G}{2}=G(B_--B_+)\\[5mm]
    s_3=F_+^{-2}\,[2F_+B_+s_1+(B_+^2-1)s_2]\\[5mm]
    G^2=s_1^2+s_2s_3.
    \end{array}
  \]
 \item Any function $\sigma\in{\cal B}_\gamma$ can uniquely be written as
 the sum \[\sigma=\sigma_++\sigma_-\] with $\sigma_+$ and
 $\sigma_-$ only possessing singularities without and within the
 curve $\Gamma$, respectively, and $\sigma_-\to 0$ as $|\,X|\to\infty$.
 The linear operators
 $\hat{L}_\pm, \hat{L}_\pm:{\cal B}_\gamma\to{\cal
 B}_\gamma$, extract these functions $\sigma_\pm$:
 \[\hat{L}_\pm\sigma=\mp\,\sigma_\pm.\]
 Explicitly:
 \[ \frac{1}{2\pi\mbox{i}}\oint_\Gamma\frac{\sigma(Y)dY}{X-Y}
 =\left\{
 \begin{array}{ll}
    (\hat{L}_+\sigma)(X)=-\sigma_+(X)&\mbox{for $X$ within $\Gamma$}\\
    (\hat{L}_-\sigma)(X)=\sigma_-(X)&\mbox{for $X$ without $\Gamma$}\\
 \end{array}
 \right.,\]
 and these functions can be extended to $\mathbb{G}_\gamma$ since
 $\sigma\in{\cal B}_\gamma$.
\end{enumerate}
By means of the substitutions
\[\begin{array} {lcl}
  F_\pm&=&1+\varepsilon F^*_\pm \\
  B_\pm&=& \varepsilon B^*_\pm,\\
  s_1&=&\varepsilon(F^*_--F^*_+)+\varepsilon^2 s_1^*\\
  s_2&=&\varepsilon(B^*_+-B^*_-)+\varepsilon^2 s_2^*\\
  s_3&=&\varepsilon(B^*_--B^*_+)+\varepsilon^2 s_3^*\end{array}\]
the differential system (\ref{Dgl}) reads
\[\frac{d}{d\varepsilon}\left(\begin{array}{l}
F_+^*\\F_-^*\\B_+^*\\B_-^*\end{array}\right)
=\left(\begin{array}{l}
h_1(\varepsilon;F_+^*,F_-^*,B_+^*,B_-^*)\\
h_2(\varepsilon;F_+^*,F_-^*,B_+^*,B_-^*)\\
h_3(\varepsilon;F_+^*,F_-^*,B_+^*,B_-^*)\\
h_4(\varepsilon;F_+^*,F_-^*,B_+^*,B_-^*)\end{array}\right)\] with
the functions
\[h_j:[0,\varepsilon_\gamma]\times{\cal B}_\gamma^{\,4}\to{\cal
B}_\gamma\] satisfying a Lipschitz condition with respect to all
arguments in a sufficiently small interval
$[0,\varepsilon_\gamma]$; the upper limit $\varepsilon_\gamma$ is
defined in this manner. Note that for the continuity of the
linear operators $\hat{L}_\pm$ it is necessary to define them on
${\cal B}_\gamma$ and not on the Banach space of functions
analytic at $\Gamma$.\\
As a consequence of the theorem by Picard and Lindel\"{o}ff, the
above system of differential equations has a solution which
depends uniquely and continuously on the given initial conditions.
This is the point at which we bring in the function $\gamma$:
\[F^*_\pm(\varepsilon=0)=\hat{L}_\pm(\gamma+\gamma^*)\,,\quad
B^*_\pm(\varepsilon=0)=-\hat{L}_\pm(\gamma-\gamma^*)\,,\] where
the function $\gamma^*\in{\cal B}_\gamma$ results from $\gamma$ by
\[\gamma^*(X)=\overline{\gamma(\overline{X})}
\quad\mbox{for $X\in\mathbb{G}_\gamma$}.\] 
These initial conditions together with 
the differential equations (\ref{Dgl}) yield 
$F_\pm$ and $B_\pm$ which only possess singularities 
without/within $\Gamma$. Thus, the functions $F$ and $B$ obey a 
Riemann-Hilbert-Problem (\ref{RHP}), and the associated functions
$\alpha=\alpha(X;\varepsilon;\gamma)$ and $\beta=\beta(X;\varepsilon;\gamma)$ 
are uniquely determined. 
Consequently, we find a formulation
$f=f(\varepsilon;\gamma)$ by solving the linear integral equation
that has been mentioned in (a) for $\alpha=\alpha(X;\varepsilon;\gamma)$ and 
$\beta=\beta(X;\varepsilon;\gamma)$.\footnote{The solubility of this
integral equation is ensured for sufficiently small
$\varepsilon$, see point 2. of the proof.}
\item If we prescribe $\gamma_p$ in the form
\[\gamma_p(X)=H_p(X)\prod_{\nu=1}^p(X-X^{(p)}_\nu)\]
then it can be shown that the Ernst potential following from the
above reads
\[f(\varepsilon;\gamma_p)=f(\{X_\nu^{(p)}\}_p\,,\varepsilon H_p).\]
The proof of this uses many of the solution methods that were
developed by Neugebauer and Meinel when they solved the boundary
value problem of the rigidly rotating disk of dust. These methods
are partially given in \cite{N00}; in full they will be treated in a
subsequent paper. \\
For any series $\{\gamma_p\}_{p_0}^\infty$ of the above functions
$\gamma_p$ with $\gamma_p\to\gamma$ as $p\to\infty$ and
$\gamma_p\in{\cal B}_\gamma$, the corresponding Ernst potentials
$f=f(\varepsilon;\gamma_p)$ converge for sufficiently small
$\varepsilon$ since the functional $f=f(\varepsilon;\gamma)$
depends continuously on $\gamma$. Because of the above equality,
this implies the convergence of
$f(\{X_\nu^{(p)}\}_p\,,\varepsilon H_p)$ as $p\to\infty$.
\end{enumerate}
{\bf Remark:} The differential system (\ref{Dgl}) can be explicitly
solved if $\gamma$ is regular without $\Gamma$ with $\gamma\to
0$ as $|\,X|\to\infty$. Then one obtains
\[\alpha=\cosh G+(\gamma+\gamma^*)\frac{\sinh G}{2G}\,,\quad
\beta=-(\gamma-\gamma^*)\frac{\sinh G}{2G}\,,\quad
G^2=\gamma\gamma^*\,,\] from which the Ernst potential along the
rotation axis can be taken directly, see (\ref{axis}) below.
\item The proof of the second statement uses the relations
$\alpha=\alpha(X;\varepsilon;\gamma)$ and
$\beta=\beta(X;\varepsilon;\gamma)$. In particular, we have
$\alpha,\beta\in {\cal B}_\gamma$ for
$\varepsilon\in[0,\varepsilon_\gamma]$ and moreover
$\alpha=1+{\cal O}(\varepsilon)$ and $\beta={\cal
O}(\varepsilon)$. This means that for sufficiently small
$\varepsilon$, the linear integral equation that yields the
Ernst potential from the functions $\alpha$ and $\beta$ can be
solved for arbitrary coordinates $(\rho,\zeta)\notin B$, see
\cite{N96}\footnote{Note that the linear integral equation might
globally only be soluble for
$\varepsilon<\varepsilon_\gamma^*<\varepsilon_\gamma$.}. Moreover,
the Ernst potential can be extended to the region
$\mathbb{B}_\gamma$ since $\alpha,\beta\in {\cal B}_\gamma$. The
different boundary values at the inner and outer sides of $B$
follow from the construction.
\item Finally, the proof of the third statement uses the linear system (\ref{LS}).
For $X\in\Gamma$ with $\Re X\geq 0$, we establish the matrices
$\tilde{\Phi}^\pm[X,\rho=0,\zeta=r_B(1)\pm0]$ by integrating the linear system along the
inner and outer side of $B$, with the initial values
\[\tilde{\Phi}^\pm(X,\Re X,\Im X)=\left(\begin{array}{ll} 1&1\\1&-1\end{array}\right).\]
The coordinate $\zeta=r_B(1)\pm0$ stands for the inner and outer
side of $B$ at $\rho=0$, respectively. The integration of the
linear system can be performed since for a sufficiently weak
relativistic regular Ernst potential again the theorem by Picard
and Lindel\"{o}ff applies. At $\rho=0$ and $\zeta=r_B(1)\pm0$, we then
establish $\Phi$ from $\tilde{\Phi}$ by $\Phi=\tilde{\Phi}M_\pm$
where the regular matrices $M_\pm$ (defined on $\Gamma$) are chosen such
that $\Phi$ assumes the structure (\ref{Pax}), i.e.~we calculate the
functions $F_\pm$ and $B_\pm$. In order to determine $\gamma$ we
choose some $\varepsilon_0>0$ (say $\varepsilon_0=1$) and
integrate the differential system (\ref{Dgl}) backwards, starting
at $\varepsilon_0$ (here the initial conditions are just the
established $F_\pm,B_\pm$), until we reach $\varepsilon=0$, and
read $\gamma$ from the weak field expansion:
\[\gamma=\lim_{\varepsilon\to 0} \frac{1}{2\varepsilon}[(F_--F_+)+(B_+-B_-)].\]
So we have proven that the interior and exterior 
boundary data $f$ and $f_{,z}\,,f_{,\bar{z}}$ 
(the derivatives also enter the linear system) of a sufficiently weak relativistic 
Ernst potential, which is regular within and without $B$, uniquely determine the function 
$\gamma$. But since the weakly relativistic 
Dirichlet boundary value problem is uniquely soluble \cite{Reula}, the global 
interior and exterior Ernst potential (and hence the above derivatives) are 
determined by the boundary values $f$ alone, which therefore determines $\gamma$ as well.
\end{enumerate}
\subsubsection*{Properties of the generalized hyperelliptic solutions.}
\begin{enumerate}
  \item For $\psi\in\mathbb{R}$ one finds 
  \begin{equation}\label{inv3} f(e^{2\mbox{\scriptsize i}\psi}\varepsilon\gamma)=
   \frac{f(\varepsilon\gamma)\cos\psi+\mbox{i}\sin\psi}
        {\cos\psi+\mbox{i}f(\varepsilon\gamma)\sin\psi}\,.\end{equation}
        This can be seen by investigating $f(\{X_\nu\}_{p+1}\,,H_R)$ as $R\to\infty$
        with
        \begin{eqnarray*}\gamma_p(X)&=&H(X)\prod\limits_{\nu=1}^p(X-X_\nu),\\
        e^{2\mbox{\scriptsize i}\psi}\gamma_p(X)
        &=&\lim_{R\to\infty}\gamma_{p+1}=
        \lim_{R\to\infty}\left[-\frac{H(X)}{R}(X-Re^{2\mbox{\scriptsize i}\psi})
        \prod_{\nu=1}^p(X-X_\nu)\right]\\&=&
        \lim_{R\to\infty}\left[H_R(X)(X-X_{p+1})\prod\limits_{\nu=1}^{p}(X-X_\nu)\right].
        \end{eqnarray*}
        Note that in the integral
        terms of (\ref{hyper}, \ref{JUP}) for $\nu=p+1$ the substitution
        $X=X_{p+1}(1+t^2)$ is useful.\\
        The property (\ref{inv3}) describes the general invariance transformation of
        $f(\varepsilon\gamma)$ which retains the asymptotic flatness
        ($f\to 1$ as $r\to\infty$). $f(e^{2\mbox{\scriptsize i}\psi}\varepsilon\gamma)$
        is obtained from $f(\varepsilon\gamma)$ as one performs the transformation (21)
        of \cite{KN68} with the parameters $\alpha=\sin^{-2}\psi,\,\beta=\gamma=-\cot\psi$ 
        of that paper.
  \item A given function $\gamma:\Gamma\to\mathbb{C}$ can be
  represented in many ways by sequences
  $\{\{X_\nu^{(p)}\}_p\}_{p_0}^\infty$ and
  $\{H_p\}_{p_0}^\infty$. Consider the following example for a spherical boundary, $r_B\equiv 1$.
  Since $\gamma$ is analytic on $\Gamma$, it can be written as
  \[\gamma(X)=\sum_{j=1}^\infty \left(\gamma_j^{(+)} X^{j-1}
    +\gamma_j^{(-)} X^{-j}\right)\,,\quad X\in\Gamma\]
  where
  \[\gamma^{(\pm)}_j=\pm\frac{2j-1}{2}\int_0^\pi
  Q_\pm(\cos\vartheta)P_{j-1}(\cos\vartheta)\sin\vartheta\,
  d\vartheta\,,\]
  \[Q_\pm(\cos\vartheta)=\lim_{r\to 1\mp 0}\left[\frac{1}{2\pi
  \mbox{i}}\oint_\Gamma\frac{\gamma(X)\,dX}{W_z}\right].\]
  For each $n\in\mathbb{N}$ we find a representation
  \[\gamma_n(X)=\sum_{j=1}^n (\gamma_j^{(+)} X^{j-1}+\gamma_j^{(-)} X^{-j})=
  R_n e^{2\mbox{\scriptsize i}\psi_n}X^{m_n}\prod_{\nu=1}^{p_n}(X-X_\nu^{(n)}).\]
  Thus we can write
  \[f(\varepsilon\gamma)=\lim_{n\to\infty}f(\varepsilon\gamma_n)=\lim_{n\to\infty}\left[
            \frac{f(\{X_\nu^{(n)}\}_{p_n}\,,\varepsilon H_n)
            \cos\psi_n+\mbox{i}\sin\psi_n}
            {\cos\psi_n+\mbox{i}f(\{X_\nu^{(n)}\}_{p_n}\,,\varepsilon H_n)
            \sin\psi_n}\right]\]
  with $H_n(X)=R_n X^{m_n}$.
\item If the analytic function
$\gamma:\Gamma\to\mathbb{C}$ is regular outside $\Gamma$ and
$\gamma\to 0$ as $|X|\to\infty$, then one finds
$f(\rho,\zeta;\varepsilon\gamma)=1$ within $B$ and
\begin{equation}\label{axis}\mbox{\normalsize $\begin{array}{lll}
f(\rho=0,\zeta>r_B(1);\varepsilon\gamma)&=&
\mbox{\Large $\left.\frac{G-s_-\sinh G}{G\cosh G+s_+\sinh G}\right|_{X=\zeta}$}\\[5mm]
f(\rho=0,\zeta<-r_B(-1);\varepsilon\gamma)&=& \mbox{\Large
$\left.\frac{G\cosh G+s_+\sinh G}{G-s_-\sinh G}\right|_{X=\zeta}$}
\end{array}$}\end{equation}
with
\[s_\pm(X)=\frac{\varepsilon}{2}\left[\gamma(X)\pm\overline{\gamma(\overline{X})}\right]\,,\quad
G^2(X)=s_+^2(X)-s_-^2(X),\quad\mbox{$X$ without $\Gamma$}.\]
Hence, in the case of a flat interior field there is an
explicit relation of the function $\gamma$ and the values
of the Ernst potential along the exterior part of the symmetry axis.
\item The Ernst potential $f=f(\varepsilon\gamma)$ possesses a reflectional
  symmetry
     \begin{equation}\label{symG}f(\rho,-\zeta;\varepsilon\gamma)=\left\{\begin{array}{lll}
       \overline{f(\rho,\zeta;\varepsilon\gamma)}\,^\kappa &\mbox{for}&\mbox{$(\rho,\zeta)$ within $B$}\\
       \overline{f(\rho,\zeta;\varepsilon\gamma)}\,^{-\kappa}&\mbox{for}& \mbox{$(\rho,\zeta)$ without $B$}
      \end{array}\right.,\end{equation}
      if $r_B(\tau)=r_B(-\tau)$ and $\overline{\gamma(-\overline{X})}=\kappa\,\gamma(X)\,,\kappa^2=1.$
      We obtain the desired reflectional symmetry of the exterior Ernst potential
      for $\kappa=-1$.
 \item There is a weak field expansion of $f=f(\varepsilon\gamma)$:
   \begin{equation}\label{weak}\ln[f(\varepsilon\,\gamma)]=\frac{\varepsilon}{2\pi
  \mbox{i}}\oint_\Gamma \frac{\gamma(X)\,dX}{W_z}+{\cal O}(\varepsilon^3).\end{equation}
  By differentiating (\ref{hyper}, \ref{JUP}) with respect to $\varepsilon$,
  where $H$ and $\{X_\nu\}_p$ are fixed,
  one deduces that
  \[\frac{d}{d\varepsilon}\ln[f(\{X_\nu\}_p\,,\varepsilon H)]=\frac{1}{2\pi
  \mbox{i}}\oint_\Gamma H(X)\prod_{\nu=1}^p(X-X^{(\nu)})\frac{dX}{W_z}.\]
  The expansion (\ref{weak}) follows since, in the weak relativistic region,
  (\ref{JUP}) yields $X^{(\nu)}=X_\nu+{\cal O}(\varepsilon^2)$.
  \item Another interesting property is 
  \[f(\varepsilon\gamma^*)=\overline{f(\varepsilon\gamma)}\quad\mbox{for}\quad
  \gamma^*(X)=\overline{\gamma(\overline{X})}.\]
  If, in particular, $\gamma=\gamma^*$ then
  $f=f(\varepsilon\gamma)$ is real and belongs to the Weyl class.
\end{enumerate}
\section{Approximation of arbitrary exterior boundary value problems by
         B\"{a}cklund type solutions}
Although the generalized hyperelliptic class permits the simultaneous solution
of a sufficiently weak relativistic exterior and interior boundary value problem of the
Ernst equation, its mathematical complexity makes it inconvenient for usage in a
procedure to approximate the solution of the boundary value problem in question. Therefore
one is led to investigate whether the much simpler B\"{a}cklund type solutions
suffice for our approximation scheme of only {\em exterior} boundary value problems.
As we will demonstrate in this section,
the freedom of the choice of the interior field allows us to set
$\gamma\neq 0$ for $X\in\Gamma$, which ensures that the corresponding Ernst potential
$f$ can be very well approximated by the B\"{a}cklund type solutions.
\subsection{The class of the B\"{a}cklund type solutions and its generalization}
\subsubsection*{Definition.}
The B\"{a}cklund type solutions $f_B(\{Y_\nu\}_q,G)$ depend on
the set $\{Y_\nu\}_q$ of complex parameters and on the analytic
function $G:\Gamma\to\mathbb{C}$ with
$\overline{G(X)}=G(\overline{X})$ and are defined by
   \begin{eqnarray}
   \label{fB}
       f_B(\rho,\zeta;\,\{Y_\nu\}_q,G)=
       f_0\frac{D_+}{D_-}
   \end{eqnarray}
   where
\[D_\pm=  \left|\begin{array}{ccccc}
     1&                1       &                 1&\cdots&                   1    \\[2mm]
\pm  1&\alpha_1\lambda_1       &\alpha_2\lambda_2 &\cdots&\alpha_{2q}\lambda_{2q} \\[2mm]
     1&\lambda_1^2             &\lambda_2^2       &\cdots&\lambda_{2q}^2          \\[2mm]
\pm  1&\alpha_1\lambda_1^3     &\alpha_2\lambda_2^3&\cdots&\alpha_{2q}\lambda_{2q}^3\\[2mm]
\vdots&                \vdots  &\vdots             &\ddots&\vdots                 \\[2mm]
\pm  1&\alpha_1\lambda_1^{2q-1}&\alpha_2\lambda_2^{2q-1}
      &\cdots&\alpha_{2q}\lambda_{2q}^{2q-1}\\[2mm]
     1&\lambda_1^{2q}        &\lambda_2^{2q}     &\cdots&\lambda_{2q}^{2q}      \\[2mm]
                    \end{array}\right|,\]
\begin{eqnarray*}
    f_0&=&\exp\left(\frac{1}{2\pi \mbox{i}}\oint_\Gamma \frac{G(X)\,dX}{W_z}\right),\\
    \lambda_{2\nu-1}&=&\sqrt{\frac{Y_\nu-\mbox{i}\bar{z}}{Y_\nu+\mbox{i}z}}\,,
     \quad \lambda_{2\nu}\,\overline{\lambda}_{2\nu-1}=1,\\
    \alpha_{2\nu-1}&=&-\tanh\left(\frac{\lambda_{2\nu-1}(Y_\nu+\mbox{i}z)}{4\pi
    \mbox{i}}\oint_\Gamma\frac{G(X)\,dX}{(X-Y_\nu)W_z}\right)\,,\\
    \alpha_{2\nu}\,\overline{\alpha}_{2\nu-1}&=&1.\end{eqnarray*}
The Ernst potential $f_B(\{Y_\nu\}_q,G)$ is only then regular
inside and outside the boundary $B$ if we additionally require
that for each $Y_\nu\in\Gamma$ there is $G(Y_\nu)=0$.\footnote{
More precisely, if $|G(Y_\nu)|>\delta>0$ always holds, then the interior/exterior Ernst
potential encounters a square root like behavior at
$(\rho,\zeta)=(|\Im[Y_\nu]|,\Re[Y_\nu])$ when $Y_\nu$ tends
to a point at the outer/inner side of $\Gamma$.} For the desired procedure for approximating
the solutions of exterior boundary value problems we will restrict ourselves to
potentials $f_B(\{Y_\nu\}_q,G)$ with all $Y_\nu\notin\Gamma$.
\subsubsection*{B\"{a}cklund type solutions as special hyperelliptic solutions.}
The B\"{a}cklund type solutions form a subclass of the hyperelliptic solutions.
In particular:
\begin{equation}\label{trans}
f_B(\{Y_\nu\}_q,G)=f(\{Y_1,Y_1,\ldots,Y_q,Y_q\}\,,H)\end{equation}
with
\[H(X)=G(X)\left[\,\prod_{\nu=1}^q(X-Y_\nu)(X-\overline{Y}_\nu)\right]^{-1}\]
and $f(\{X_\nu\}_p,H)$ as defined in (\ref{hyper}). The proof of
this works in the same manner as demonstrated in \cite{A01} for the
solutions corresponding to disk like sources. The regularity of
the solutions $f(\{X_\nu\}_p,H)$ at
$(\rho,\zeta)=(|\Im[X_\nu]|,\Re[X_\nu])$ applies in this
specialization and is exhibited by the fact that $\alpha_j\lambda_j$ is an even
function in $\lambda_j$. This means that $f$ does not behave like
a square root function near the critical points
$(\rho,\zeta)=(|\Im[Y_\nu]|,\Re[Y_\nu])$ but rather like a
rational function.
\subsubsection*{B\"{a}cklund transformations.}
The Ernst potentials (\ref{fB}) are special B\"{a}cklund
transformations as described in \cite{N96} (see formula 76) with complex
conjugate parameters $K_{2\nu-1}=Y_\nu,\,
K_{2\nu}=\overline{Y}_\nu$ and the real seed solution
$f_0=\exp[U(G)]$. The $\alpha_j$ in 
formula (76) of \cite{N96} satisfy the Riccati equations 
\[\alpha_{j,z}=\lambda_j(1-\alpha_j^2)\frac{f_{0,z}}{2f_0},\quad
  \alpha_{j,\bar{z}}=\frac{1}{\lambda_j}(1-\alpha_j^2)
  \frac{f_{0,\bar{z}}}{2f_0}. \]
With constants $C_j$ of integration, the general solution of
these equations reads
\[\alpha_j=-\tanh\left(\frac{\lambda_j(K_j+\mbox{i}z)}{4\pi
    \mbox{i}}\oint_\Gamma\frac{G(X)\,dX}{(X-K_j)W_z}+C_j\right)\,.\]
By the particular choice in (\ref{fB}) 
[i.e.~$C_{2\nu-1}=0\,,\,C_{2\nu}=\mbox{i}\pi/2$], the
$\alpha_j$ become odd functions in $\lambda_j$ which ensures the
regularity of the resulting Ernst potential at
$(\rho,\zeta)=(|\Im[Y_\nu]|,\Re[Y_\nu])$.
\subsubsection*{The generalization of the B\"{a}cklund type solutions.}
The function $\gamma_q$ which belongs to the Ernst potential $f=f_B(\{Y_\nu\}_q,G)$
can be taken from (\ref{trans}):
\[\gamma_q(X)=G(X)\prod_{\nu=1}^q\frac{X-Y_\nu}{X-\overline{Y}_\nu}.\]
It is therefore a consequence of the above theorem that for given
analytic functions $G:\Gamma\to\mathbb{C}$ and
$\Xi:\Gamma\to\mathbb{C}$ with $\overline{G(X)}=G(\overline{X})$
and $\overline{\Xi(X)}=-\Xi(\overline{X})$, as well as sufficiently small $\varepsilon$,
the Ernst potential
\[f_B(\Xi,\varepsilon G)=\lim_{q\to\infty}f_B(\{Y_\nu^{(q)}\}_q,\varepsilon G)\]
exists and is independent of the particular choice of the sequence
$\{\{Y_\nu^{(q)}\}_q\}_{q_0}^\infty$ which serves to represent
$\Xi$ by
\begin{eqnarray}\label{G2.4}
\Xi(X)=\lim_{q\to\infty}\,
     \ln\left[\,\prod_{\nu=1}^q\frac{X-Y_\nu^{(q)}}{X-\overline{Y}_\nu^{(q)}}\right]
\quad \mbox{ for}\quad X\in\mathbb{G}_G\cap\mathbb{G}_\Xi.    \nonumber
\end{eqnarray}
In particular we have
\begin{equation}\label{fBf}
f_B(\Xi,\varepsilon G)=f(\varepsilon \gamma) \quad\mbox{with}\quad
\gamma(X)=G(X)\exp[\Xi(X)].\end{equation} From (\ref{fBf}) we may
conclude that any analytic function $\gamma$ which possesses a
single zero $X_\nu\in\Gamma$ and does not vanish at
$\overline{X}_\nu$ cannot be represented by two analytic
functions $G$ and $\Xi$. 
Therefore, in the following we will restrict the discussion to $\gamma\neq 0$
for $X\in\Gamma$.\\[5mm]
 {\bf Remark: } As a particular
consequence we now can confirm the conjectures formulated in
\cite{A01}, sections 2 and 5. The first one of these conjectures treats
the generalization of B\"{a}cklund type solutions that describe
disk-like sources of the gravitational field. In order to map the
treatment of that place to our extended $B$ here, we consider 
for given functions $\xi$ and $g$ \footnote{In \cite{A01}, $\xi$ and $g$
were assumed to be real analytic functions defined on the interval $[0,1]$.
$g$ describes the seed solution while $\xi$ comprises the
B\"{a}cklund parameters.} and an arbitrary point 
$(\rho,\zeta)$ without the disk, a boundary
$B$ which encompasses the disk but leaves the point
$(\rho,\zeta)$ outside $B$ as well as all singularities of 
$g(-X^2/\rho_0^2)$ and $\xi(-X^2/\rho_0^2)$
outside $\Gamma$ ($\rho_0$ is the radius of the disk). 
If we now take any sequence $\{\{Y_\nu^{(q)}\}_q\}_{q_0}^\infty$ 
which serves to represent $\xi$ by
\begin{eqnarray*}
\xi(x^2)=\lim_{q\to\infty}\xi_q(x^2)\,,\quad\xi_q(x^2)=\frac{1}{x}
     \ln\left[\,\prod_{\nu=1}^q\frac{\mbox{i}\,Y_\nu^{(q)}-x}{\mbox{i}\,Y_\nu^{(q)}+x}\right]
\quad \mbox{ for}\quad x\in[-1,1]    \nonumber
\end{eqnarray*}
with all $(\rho_0 Y_\nu^{(q)})$ being outside $\Gamma$, then the 
resulting Ernst potential $f(\rho,\zeta;\xi_q,g)$ in \cite{A01} at the chosen point $(\rho,\zeta)$ 
coincides with the above $f_B(\rho,\zeta;\Xi_q,G_q)$ when
\begin{eqnarray*}G_q(\mbox{i}\rho_0
x)&=&2(-1)^{q+1}\mbox{arccot}(\mbox{i}x)g(x^2)\,,\\
\exp[\Xi_q(\mbox{i}\rho_0 x)]&=&(-1)^q\exp[-x\xi_q(x^2)]\,.\end{eqnarray*}
As $q\to\infty$, the corresponding functions $\gamma_q$ with
\[\gamma_q(\mbox{i}\rho_0 x)=G_q(\mbox{i}\rho_0 x)\exp[\Xi_q(\mbox{i}\rho_0 x)]\]
converge towards the function $\gamma$,
\begin{equation}\label{gxiga}
\gamma(\mbox{i}\rho_0 x)=-2\,\mbox{arccot}(\mbox{i}x)\,g(x^2)\exp[-x\xi(x^2)].
\end{equation}
So, due to the above theorem, the Ernst potentials $f(\rho,\zeta;\xi_q,g)$ in 
\cite{A01} tend to $f(\rho,\zeta;\gamma)$.
Hence we can conclude, that the exterior gravitational
field of a disk-like source is determined only by the local
functions $g$ and $\xi\,$, and does not
depend on a particular global representation of $\xi$ in terms of the
parameters $Y_\nu$. This is just the statement of the assumption
in section 2. of \cite{A01}. Note that the condition that $Y_\nu$ be
outside the imaginary interval $[-\mbox{i},\mbox{i}]$ is always
realized for sufficiently weak relativistic differentially
rotating disks of dust since this is ensured by the positivity of
the surface mass density $\sigma_p$ (see appendix C.1.2 of \cite{A01};
the function $g_0$ is strictly positive and therefore the
function $\xi_0$ is analytic). As one moves to more
relativistic disks one might encounter a situation in which $g$
and $\xi$ are not analytic (corresponding to the situation
discussed above). However, for the examples investigated, this
situation did not occur.\\
In section 5.~of \cite{A01}, conjectures were given regarding the functions $g$ 
and $\xi$ for the disk-like hyperelliptic class of solutions. These conjectures 
are also proven by the above considerations. A specially chosen 
disk-like hyperelliptic solution as presented in appendix A of \cite{A01}
with an analytic function $h$ and parameters $\{X_\nu\}_p$, 
coincides with $f(\rho,\zeta;\gamma)$ at a given point $(\rho,\zeta)$ 
outside the disk, when $\gamma$ reads 
\[\gamma(\mbox{i}\rho_0 x)=-2\,\mbox{arccot}(\mbox{i}x)\gamma_D(\mbox{i}\rho_0 x) \,,
\quad\gamma_D(\mbox{i}\rho_0 x)=h(x^2)\prod_{\nu=1}^p (\mbox{i}x-X_\nu)\]
and a corresponding boundary $B$ is taken in the above 
manner with the singularities of $\gamma_D$ being outside $\Gamma$. 
The comparison with (\ref{gxiga}) yields  exactly the conjectured functions
$g$ and $\xi$ in terms of $h$ and $\{X_\nu\}_p$. 
It is of particular interest that thus the
Neugebauer-Meinel-solution can be written as a well defined limit of 
B\"{a}cklund type solutions. 
\subsection{The approximation scheme and results}
In this section we describe our procedure to approximate the solution to an
exterior boundary value problem of the Ernst equation.
Due to the preceding formulations, we seek
an appropriate function $\gamma$ such that the corresponding Ernst potential
$f=f(\gamma)$ assumes prescribed exterior boundary values at $B$.
As outlined above, we only consider functions $\gamma$ that do not vanish at $\Gamma$, 
and for which the corresponding Ernst potentials can be well approximated by the B\"{a}cklund
type solutions. We find in particular that in contrast to the Laplace equation, arbitrary
boundary values of the Dirichlet type are not possible.
\subsubsection*{The approximation scheme.}
\begin{enumerate}
\item At first we restrict the function $\gamma$ in an apropriate manner. Because of the
      desired reflectional symmetry (\ref{symG}) we take $r_B(\tau)=r_B(-\tau)$ and
      $\overline{\gamma(-\overline{X})}=-\gamma(X)$.
      To avoid zeros at $\Gamma$ we demand
      $\Im(\gamma)=\mbox{const.}=\gamma_0\neq 0$ for $X\in\Gamma$. This leads to
      \[\gamma(X)=\mbox{i}\gamma_0+A_1(X)+A_2(X)\]
      with
      \begin{eqnarray*}A_1\left[X=r_B(\cos\theta) e^{\mbox{\scriptsize i}\theta}\right]
         &=&\sum_{k=1}^\infty\gamma_{2k-1}\cos[(2k-1)\theta]\,,\\
      A_2\left[X=r_B(\cos\theta) e^{\mbox{\scriptsize i}\theta}\right]
         &=&\sum_{k=1}^\infty\gamma_{2k}\sin[2k\theta]
        \,,\quad\gamma_j\in\mathbb{R}\,,\quad\theta\in[0,2\pi].\end{eqnarray*}
      The real constant $\gamma_0\neq 0$ may be chosen arbitrarily; the coefficients
      $\gamma_j$ for $j\geq 1$ are then determined uniquely by the exterior boundary
      conditions.
\item It is now possible to find analytic functions $\hat{G}$ and $\hat{\Xi}$ with
\begin{eqnarray*}
\gamma(X)&=&\mbox{i}\,\hat{\gamma}(X)=
           \mbox{i}\,\hat{G}(X)\exp[\hat{\Xi}(X)],  \\
\hat{G}(X)&=&\sqrt{A_1^2(X)-\left[A_2(X)
+\mbox{i}\gamma_0\right]^2}\,,\quad\hat{G}\left[X=\mbox{i}\,
r_B(0)\right]=\gamma_0,\\
\hat{\Xi}(X)&=&-\frac{\mbox{i}}{2}\left[\arctan\left(\frac{A_1(X)+A_2(X)}{\gamma_0}\right)
       +\arctan\left(\frac{A_1(X)-A_2(X)}{\gamma_0}\right)\right] \\& &
+\frac{1}{4}\ln\left[\frac{\left[A_1(X)+A_2(X)\right]^2+\gamma_0^2}
                                {\left[A_1(X)-A_2(X)\right]^2+\gamma_0^2}\right],\quad
\hat{\Xi}\left[X=\mbox{i}\,
r_B(0)\right]=0.
  \end{eqnarray*}
       With (\ref{inv3}), the Ernst potential reads then as follows
\[f(\gamma)=\frac{f_B(\hat{\Xi},\hat{G})+\mbox{i}}
 {1+\mbox{i}f_B(\hat{\Xi},\hat{G})}.\]
\item For a given analytic function $\hat{\Xi}$ of the above kind we determine
      corresponding B\"{a}cklund parameters $\{Y_\nu\}_q$ in the same
      manner as outlined in \cite{A01}. This means that we solve the
      linear system:
      \[\exp[\hat{\Xi}(Z_\nu)]P(-Z_\nu)=(-1)^q P(Z_\nu),\quad\nu=1\ldots q,\]
      \[P(X)=\prod_{\nu=1}^q(X-Y_\nu)=X^q+\sum_{j=0}^{q-1}b_jX^j\]
      to determine the coefficients $b_j$ and from these the zeros
      $Y_\nu$ of $P$. The $q$ arbitrary different supporting points
      $Z_\nu\in\Gamma$ are chosen to possess positive real and imaginary parts.
\item By virtue of formula (\ref{fB}) we are now able to approximate the Ernst potential
      $f(\gamma)$ if the parameters $\{\gamma_j\}_0^\infty$ are given. For the
      approximate numerical evaluation of $\{\gamma_j\}_1^{2n}$
      from the given exterior boundary values we set $\gamma_j=0$ for $j>2n$ and demand that
      the resulting exterior Ernst potential $f(\gamma)$ coincide with the prescribed
      boundary values at $n$ different points $(\rho_j,\zeta_j)\in B$ with
      $\zeta_j\geq 0$. This gives a complicated nonlinear set of $2n$ real equations
      to determine the unknown coefficients $\{\gamma_j\}_1^{2n}$.
      As in the analogous treatment to solve boundary value problems for arbitrary
      differentially rotating disks of dust, we solve this system by means of a
      Newton-Raphson method. Again we provide good initial guesses by solving a sequence
      of boundary value problems with initially weak relativistic and
      finally the desired boundary values, and where the initial guess comes from
      the weak relativistic expansion (\ref{weak}).
\end{enumerate}
\subsubsection*{Results.}
The above approximation scheme has been executed for various Dirichlet boundary value
problems. As expected from the above, there is always a weak relativistic region
within which the solution exists and can be well approximated by the
B\"{a}cklund type solutions. In particular, the exterior solution proves to be independent
of the choice of the parameter $\gamma_0$, only the accuracy of the results is
affected. \\
As one moves further to more relativistic boundary values,
there might be however a limit beyond which the solutions cannot be extended. For example,
take the exterior boundary values
\[f\left[(\rho,\zeta)\in B\right]=1+\varepsilon\left[-\frac{3}{2}+\zeta^2\right]+\mbox{i}\varepsilon\zeta,\]
with the spherical boundary $B$ characterized by $r_B\equiv 1$. One finds a
limiting parameter $\varepsilon_0\approx 0.68$ beyond which the
boundary value problem does not seem to be soluble. This has
been verified not only by various choices of the above parameter
$\gamma_0$ but also by a two-dimensional numerical method, which
solves the exterior Ernst equation directly without using analytic
solutions\footnote{For this method, we insert a two-dimensional
Chebyshev expansion of the exterior Ernst potential into the
Ernst equation and find the Chebyshev coefficients.}. All these
routines yield the {\it same} limiting parameter $\varepsilon_0$
which is a strong indication that for $\varepsilon>\varepsilon_0$ the
Ernst potentials corresponding to the above boundary values do not exist.

\section*{Acknowledgement}
The support from the DFG is gratefully acknowledged.

\end{document}